\documentclass[
10pt,tightenlines, amsmath, amssymb, nofootinbib, prd,
superscriptaddress, showpacs, preprintnumbers]{revtex4}

\newcommand{\beq}{\begin{equation}}
\newcommand{\eeq}{\end{equation}}
\newcommand{\bea}{\begin{eqnarray}}
\newcommand{\eea}{\end{eqnarray}}
\newcommand{\bmat}{\begin{pmatrix}}
\newcommand{\emat}{\end{pmatrix}}

\def\d{\partial}
\def\<{\langle}
\def\>{\rangle}
\def\+{\dagger}

\def\dm{{\partial}_{\mu}}
\def\eps{\epsilon^{\mu \nu \lambda \sigma}}
\usepackage{graphicx}
\usepackage{amsmath}
\usepackage{amssymb}
\usepackage{psfrag}
\usepackage{epsfig}
\usepackage{graphicx}
\usepackage{epstopdf}
\begin{document}
\preprint{BNL-NT-07/24}
\title{Charge separation induced by ${\cal P}$--odd bubbles in QCD matter}
 
\author{Dmitri~Kharzeev}
\email{kharzeev@bnl.gov} \affiliation{Physics Department, Brookhaven National Laboratory, Upton, NY 11973-5000, USA}
 
\author{Ariel~Zhitnitsky}
\email{arz@physics.ubc.ca} \affiliation{Department of Physics and
Astronomy, University of British Columbia, Vancouver, BC, Canada,
V6T 1Z1}

\date{\today}
\begin{abstract}
We examine the recent suggestion that ${\cal P}-$ and ${\cal CP}-$odd effects in QCD matter 
can induce electric charge asymmetry with respect to reaction plane in relativistic heavy ion collisions. General arguments are given which confirm that the angular momentum of QCD matter 
in the presence of non--zero topological charge should induce an electric field aligned along the axis 
of the angular momentum. A simple formula relating the magnitude of charge asymmetry to the angular momentum and topological charge is derived. The expected asymmetry is amenable to experimental observation at RHIC and LHC; we discuss the recent preliminary STAR result in light of our findings. Possible implications of charge separation phenomenon in cosmology and astrophysics are discussed as well. 
\end{abstract}

\maketitle
\section{Introduction}

The strong ${\cal CP}$ problem remains one of the most outstanding puzzles of the Standard Model. 
The best known solution of this problem is provided by the axion, see the original papers \cite{axion},\cite{KSVZ},\cite{DFSZ},
and recent  reviews \cite{review}. Thirty years after the axion was invented, it is still considered as
a viable solution of the strong ${\cal CP}$ problem, and 
one of the plausible dark matter candidates. At present,  there are several experimental groups
 searching for the axions, see 
\cite{vanBibber:2006rb} and references therein.
 The axion scenario 
suggests that in the Early Universe ${\cal P}$ and ${\cal CP}$ invariances were strongly violated, while
at present time they are preserved by strong interactions. ${\cal P}$ and ${\cal CP}$ conservation in 
strong interactions may thus be related to the vacuum structure. 
Moreover, recent progress in understanding the vacuum structure of gauge theories, especially supersymmetric ones \cite{SUSY1,SUSY2,SUSY3}, points to the possible existence of ${\cal P}$ and ${\cal CP}$--odd vacuum states. It was proposed that QCD vacuum in the vicinity of the deconfinement phase transition can also possess ${\cal P}-$ and ${\cal CP}-$odd metastable vacuum solutions \cite{KPT} possibly inducing ${\cal CP}$--odd correlations of the produced hadrons \cite{KP} in relativistic heavy ion collisions.
Such "${\cal P}$--odd bubbles" are a particular realization of an excited vacuum domain which may be produced in heavy ion collisions 
\cite{LeeWick}, and several other realizations have been proposed before \cite{Morley,DCC}. 

\medskip

The study of metastable ${\cal P}$ and ${\cal CP}$--odd domains at RHIC 
provides a unique link to  
 the QCD phase transition in the Early Universe when $\theta$ parameter was likely not zero.
Such a large ${\cal CP}$   violation during the QCD phase transition might lead 
to a separation of matter and anti-matter in the Early Universe.
  As is well known,
 all terrestrial experiments (conducted when $\theta$ has already relaxed to zero,  
 as the axion solution of the strong ${\cal CP}$ problem suggests)
 show very tiny ${\cal CP}$ violation which is not sufficient to explain the baryon number density observed today. 
 We speculate in Section IV on a number of other  astrophysical/cosmological implications which nonzero $\theta$ in a few moments after the Big Bang might have. 
 
\medskip

Several dynamical scenarios for the decay of ${\cal P}$--odd bubbles have been considered \cite{decay}, and numerical lattice calculations of 
 the fluctuations of topological charge in classical Yang--Mills fields have been performed \cite{Raju,Tuomas}. A closely related topic is the role of topological effects in high--energy scattering and 
 multi--particle production which has been studied in Refs \cite{Moch:1996bs},\cite{Kharzeev:2000ef},\cite{Shuryak:2000df},\cite{Nowak:2000de},\cite{Janik:2002nk}.
 The studies of ${\cal P}$-- and ${\cal CP}$--odd correlations of pion momenta \cite{Voloshin,Sandweiss}, including those proposed in ref\cite{miklos}, have shown that such measurements are in principle feasible but would require large event samples. 

 \section{Charge Separation Effect}

\medskip

Recently, it was proposed that ${\cal P}$ and ${\cal CP}$ violation should manifest itself 
in heavy ion collisions through the separation of electric charge with respect to the reaction plane \cite{Kharzeev:2004ey}. 
This phenomenon is a consequence of topological charge fluctuations in the presence of 
non--zero angular momentum (and/or magnetic field) in off-central heavy ion collisions. The experimental detection of this effect is possible by the method proposed in ref \cite{Voloshin:2004vk} which correlates the emission angles of charged particles with the reaction plane on the event-by-event basis. First preliminary result of such a study has been 
presented recently  by STAR Collaboration at RHIC \cite{Selyuzhenkov:2005xa}. The goal of this Letter is to re--examine the charge separation effect from a different point of view, to study the dynamics of this phenomenon in more detail and to provide numerical estimates needed for the interpretation of experimental results.

\medskip

Let us start by considering the effects of angular momentum on QCD matter. In experiment, 
hot QCD matter is produced by colliding relativistic heavy ion collisions, and such collisions occur at 
a finite impact parameter $b$. The initial angular momentum of the system is classically given by $\sim b \sqrt{s}$  and is huge at high energies $\sqrt{s}$; much of it is carried away by spectator nucleons but the produced hot matter nevertheless possesses significant angular momentum which combined with the gradients of pressure leads to the azimuthal anisotropy of hadron production. In the absence of parity violation, the angular distribution of the produced hadrons is symmetric with respect to the angular momentum axis (or equivalently, about the reaction plane) which of course fluctuates event-by-event. When expanded in harmonics, this angular distribution possesses contributions from even (as required by parity for a symmetric around zero rapidity interval) orbital momenta $l$. The harmonics of at least up to $l = 4$ have been reliably identified in the data -- see e.g.  \cite{Bai:2007ky} and references therein.
 
 \medskip
 
What are the effects of the angular momentum on the constituents of matter? Larmor theorem 
states that classical angular precession with frequency $\Omega$ for a particle of mass $m$ and electric charge $e$ is equivalent to the motion in magnetic field of strength
\beq\label{corr}
\vec{B} = {2m \over e}\ \vec{\Omega}.
\eeq 
Even more important for what follows is the fact that this correspondence persists on the quantum level; in particular, rotation of Dirac particle along the loop bounding an area $\Sigma$ yields the phase shift 
of the wave function given by 
\beq
\label{omega}
\Delta \Phi_{rot} = 2m \int \vec{\Omega} \ d\vec{\Sigma},
\eeq 
equivalent to the Aharonov-Bohm phase acquired by the motion in magnetic field $B$ threading the same loop:
\beq\label{flux}
\Delta \Phi_{B} = e \int \vec{B} \ d\vec{\Sigma}.
\eeq
This phenomenon is known as Sagnac effect; it has been firmly established experimentally 
(see refs \cite{sagnac} for reviews). Having the correspondence (\ref{corr}) in mind, we will throughout the paper substitute the influence of angular momentum by the action of a background magnetic field. 
 Such a substitution is natural also because off-central heavy ion collisions create magnetic fields aligned perpendicular to the reaction plane in the collision region occupied by hot matter. 
 \medskip
 
Let us assume, following the proposals outlined above, that a dynamical local fluctuation of $\theta$--angle can be excited in QCD matter. 
From the viewpoint of the effective lagrangian, $\theta(x)$ is equivalent to a pseudo-scalar flavor-singlet quark--anti-quark field which couples to electromagnetism through the electric charges of its quark constituents 
as prescribed by axial anomaly: 
\beq
\label{E4}
L=\frac{1}{2} \vec{E}^2 - \frac{1}{2} \vec{B}^2 +
  N_c \sum_f \frac{e_f^2}{4 \pi^2}\cdot    \left(\frac{\theta}{N_f}\right)  \left(\vec{E}\cdot \vec{B}\right),
 \eeq
 where the sum runs over quark flavors $f$, and $N_c$ is the number of colors.
Let us now minimize the action density (\ref{E4}) with respect to the electric field $E$:
\beq
\label{EB}
\frac{\delta L}{\delta E}= \vec{E}+
   N_c \sum_f \frac{e_f^2}{4 \pi^2}\cdot   \left(\frac{\theta}{N_f} \right) \vec{B} =0;
 \eeq
 we see that the magnetic field in the presence of $\theta \neq 0$ generates an electric field $E \sim \theta  \cdot B$. 
 
 \medskip
 
 To understand this phenomenon better, we start by considering an example of a field $B$ 
 provided by a magnetic monopole of magnetic charge $g$:  $  \vec{B}=({g}/{r^2})\ \vec{n}$. Using in (\ref{EB}) $N_f =1, N_c =1, e_f = e$ as appropriate in this Abelian case, we get
\beq 
\label{monopole}
\vec{E}=-\vec{n}\cdot  \frac{1}{r^2}\cdot 
   \left(\frac{e \cdot g }{2\pi}\right)\cdot \left( \frac{e \theta}{2\pi}\right), ~~ \vec{n}\equiv \frac{\vec{r}}{r}.
\eeq
Replacing $ (e \cdot g )/{2\pi}=1$ (or in general an integer) in accord with Dirac's quantization condition in eq. (\ref{monopole})
we reproduce the famous result by Witten \cite{Witten:1979ey}: magnetic monopole in the presence of $\theta$ receives the electric charge equal to $ - (e\ \theta) / 2\pi$ and becomes a "dyon" \cite{Zwanziger:1968rs}, \cite{Goldhaber:1976dp}, \cite{Wilczek:1981dr}. In fact, dyons in QCD have been known to induce chirality violation and were proposed as a solution to the $U_A(1)$ problem \cite{Pagels:1975qt}, \cite{Marciano:1976as}. A related phenomenon in grand unified theories is the proton decay induced by a magnetic monopole  -- Callan-Rubakov effect  \cite{Callan:1982ah}, \cite{Rubakov:1982fp}.

\medskip

 Let  us now consider the case of a uniform magnetic field  $B_z$ pointing in the $z$ direction; we will 
 assume that the field does not depend 
 on $x$ and $y$ coordinates, so we are dealing with an effectively 2-dimensional theory. Note that according to (\ref{corr}) this is the field which  is 
 equivalent to the action of angular momentum aligned along $z$. 
 In this case, because of the quantization of the flux (\ref{flux}) required by the single-valuedness of the particle wave function, we substitute  
  $  \int d^2x_{\perp}
 {B_z}=\Phi/e= 2\pi l /e$ into (\ref{EB}) ; $l$ is an integer.
We thus get (taking $N_f=1, e_a=e$ ),
\beq 
\label{2d}
L^2 E_z= -  \left( \frac{e \ \theta}{2\pi}\right) l,
\eeq
where $L$ is the size of the system. 
  Therefore, the electric field along $z$ will be induced in the presence of
nonzero $\theta\neq 0$ when
magnetic field is applied along  the $z$ direction. This is essentially the generalization of Witten's result \cite{Witten:1979ey} for a non-monopole like magnetic field.  The situation here resembles the  $2d$ Schwinger model with $\theta\neq 0$ \cite{Coleman:1975pw} -- indeed, this theory can be considered as occurring in a background electric field of strength $ (e\ \theta) / 2\pi$ \cite{Wilczek:1981dr}.

\medskip

Now we want to use the analogy    (\ref{corr}) to argue  that an electric field
will be induced even when 
an external magnetic field  is zero $\vec{B}\equiv 0$, but instead the system is rotating with an angular velocity
$\vec{\Omega}$. In other words we anticipate a relation $\vec{E}\sim   \left( \frac{e \ \theta}{2\pi}\right) \vec{\Omega}$ which replaces eq. (\ref{EB}) when the magnetic field is replaced by $\vec{\Omega}$.
Amazingly, such an exact relation indeed can be derived using the anomalous effective lagrangian
approach \cite{SZ} in the presence of matter with nonzero chemical potential $\mu$, and its generalization to nonzero $\theta$  ~\cite{MZ}.   

 One can show that 
the electric field will be induced for the rotating system  $\vec{\Omega}\neq 0$
according to the following relation, see Appendix for details.
\beq
\label{E}
\frac{\delta L}{\delta E}= \vec{E}+
N_c  \sum_f \frac{e_f \mu_f}{ 2\pi^2}\cdot      \left(\frac{\theta}{N_f} \right)  \cdot  \vec{\Omega} =0,
~~~~ \vec{E}=-N_c  \sum_f \frac{e_f \mu_f}{ 2\pi^2 }\cdot      \left(\frac{\theta}{N_f} \right)  \cdot {  \vec{\Omega}},
 \eeq
 where $\mu_f$ is the chemical potential of the species $f$.
 This result is quite remarkable -- it indicates that the angular momentum     
 in the presence of $\theta \neq 0$ generates a background electric field even in the absence of magnetic field. 

The electric field (\ref{E}) aligned along the vector of angular momentum (and thus perpendicular 
to the reaction plane) will act on charged particles and cause the electric charge separation signaling 
the violation of $\cal{P}$ and $\cal{CP}$ invariances arising from $\theta \neq 0$. One can  now 
easily estimate the magnitude of the expected effect once the crucial ingredient (\ref{E}) is known. 
A detailed modeling of the charge separation phenomenon would require the knowledge of charge transport properties, the geometry of the system, etc. 
Fortunately, we do not have to do all these additional assumptions because an exact formula
(for charge density
   accumulated at  the opposite sides $z=\pm L/2$ of the system of size $L$) can be derived, see Appendix,
\beq
\label{charge}
\sigma_{xy}\equiv \frac{Q}{L_xL_y}= 
  -N_c\sum_f \frac{e_f \mu_f  }{ 2\pi^2}\cdot    \Omega_z \frac{\theta}{N_f}, 
 \eeq
 where we assume that the system containing $\theta \neq 0$ domains ($\cal{P}-$ odd bubbles) has sufficient time to equilibrate\footnote{Some arguments indeed do  suggest that the system has sufficient time for the  $\theta$ vacua to be formed   at RHIC \cite{decay}.}.
 
 This result again indicates that the angular momentum     
 in the presence of $\theta \neq 0$ generates a background electric field even in the absence of magnetic field. Moreover, the magnitude of the charge density does not depend on the size
 of the of the system along $z$ direction. Of course, this result is a consequence of
 topological (anomalous) nature of the phenomenon. 
 Now the result (\ref{charge}) is very easy to interpret: the induced charge separation 
 (\ref{charge})  will result in induced $z$ independent electric field (\ref{E})   as long as 
 the system can be considered effectively two-dimensional, $L_x, L_y\gg L_z$. 
  As is well known, the electric field $E_z$  between two 
infinitely large  charged plates   with charge density $\sigma_{xy}$ is equal 
$E_z=\sigma_{xy}$. Our equations (\ref{E}) and (\ref{charge})  
satisfy this equation for an infinitely large capacitor.

\section{Numerical estimates}
 We now want to get from (\ref{charge}) an estimation for an excess of quarks over antiquarks, say, in the upper hemisphere, i.e. above the reaction plane (the lower hemisphere will obviously have an equal excess of antiquarks over quarks).   Therefore, integrating (\ref{charge}), see eq.(\ref{6}) in Appendix, 
and assuming  $\mu_u=\mu_d $ we have
   \beq\label{est1}
 Q=\int \sigma_{xy}d \Sigma_{xy}
 = -N_c\sum_f \frac{e_f   }{ \pi}   \frac{\theta}{N_f} l.
 \eeq
  For a numerical estimate we put
 $ N_c=3, N_f=2$ to arrive at 
  \beq\label{est2}
 N_{q-\bar{q}} = 2Q\simeq 3 e\left(\frac{\theta}{\pi}\right) l,
 \eeq
 which shows that the asymmetry vanishes for strictly central collisions when the angular momentum is equal to zero, $l=0$ and increases for more peripheral collisions. Note also that  
this formula and Eq.(\ref{2d}) clearly resemble  the  $2d$ Schwinger model  
with $\theta\neq 0$, see \cite{Wilczek:1981dr}.  

\begin{figure}[t]
\begin{center}
 \includegraphics[width = 0.45\textwidth]{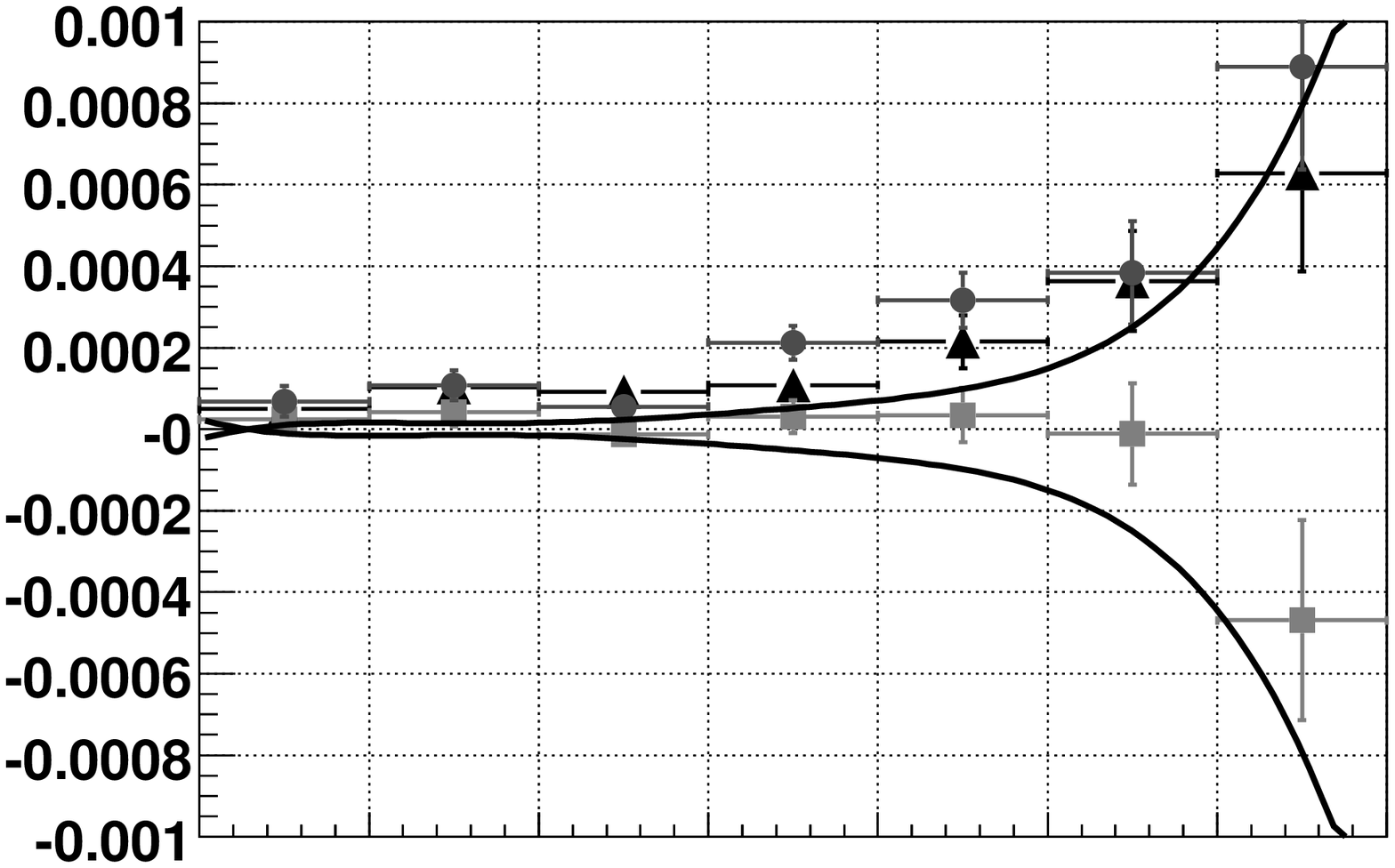}
 \put(-130,-8){$\sigma/\sigma_{tot}$, \%}
\put(-201, 7){\tiny\bf 0-5\%}
\put(-177, 7){\tiny\bf 5-10\%}
\put(-150, 7){\tiny\bf 10-20\%}
\put(-121, 7){\tiny\bf 20-30\%}
\put(-93, 7){\tiny\bf 30-40\%}
\put(-65, 7){\tiny\bf 40-50\%}
\put(-35, 7){\tiny\bf 50-60\%}
\put(-260,50){\rotatebox{90}{$a_{\pm}^2~,~~ a_{+}a_{-}$}}
\caption{\label{parityCheck}
Preliminary STAR data from \cite{Selyuzhenkov:2005xa}.
Charged particle asymmetry parameters as a a function of centrality bins
selected on the basis of charged particle multiplicity in $|\eta| < 0.5$ region.
Points are STAR preliminary data for Au+Au at $\sqrt{s_{NN}} = 62$ GeV:
circles are $a_{+}^2$, triangles are $a_{-}^2$ and squares are $a_{+}a_{-}$, see text for details.
Black lines are theoretical prediction \cite{Kharzeev:2004ey} corresponding to the topological charge $|Q|=1$.
}
\end{center}
\end{figure}
\medskip
In the presence of a collective expansion of the produced system (as evidenced by the present RHIC data), the spatial separation of charges will result in the momentum anisotropy of the emission of positive and negative particles. 
Depending on the sign of $\theta$ and $l$ which fluctuate on the event-by-event basis, an upper hemisphere can thus have either excess of quarks over anti-quarks or viceversa.  
 Taking for the sake of estimate $\frac{\theta}{\pi}\sim 1$, using $l = 4$ for a semi-central $Au-Au$ event as identified at RHIC \cite{Bai:2007ky}, and $e = \sqrt{4 \pi \alpha_{em}}$, we find that a typical event at RHIC would have on the order of 5 more quarks than antiquarks in the upper hemisphere (above the reaction plane), and an equal excess of antiquarks over quarks in the lower hemisphere. Assuming, in accord with the local parton-hadron duality, that the multiplicity of quarks and antiquarks at hadronization is approximately equal to the multiplicity of produced hadrons\footnote{It is assumed that the produced gluons split into quark--antiquark pairs before they hadronize.} which for semi-central collisions at RHIC is ${\cal{O}}(500)$, we find an asymmetry 
between quarks and antiquarks
\beq
A \equiv { N_{q-\bar{q}} \over N_{q + \bar{q}} } \sim {\cal{O}}(3 \div 5 \%),
\eeq
which agrees with the estimate of ref  \cite{Kharzeev:2004ey}. 
The effect of this magnitude may be amenable to experimental observation; a precise way for extracting 
the charge asymmetry is offered by the mixed harmonics method as proposed by Voloshin \cite{Voloshin:2004vk}. In this approach, 
one studies the correlation of the produced particles with the reaction plane on the event-by-event basis. 
The first preliminary studies of charge separation effect have been performed by the STAR Collaboration   \cite{Selyuzhenkov:2005xa} at RHIC\footnote{ We stress the preliminary nature of the reported data and the articulated by the authors of ref \cite{Selyuzhenkov:2005xa} necessity to analyze all possible systematic errors.}. 
The parameters $a_{\pm}$ plotted in Fig.1 in the absence of directed flow (which should be the case for a rapidity interval symmetric around rapidity $y=0$ in a symmetric collision) can be represented as 
\beq\label{defa}
a_i a_j = \left< \sum_{k,l}\ \sin(\phi^k_i - \psi_{RP}) \sin(\phi^l_j - \psi_{RP}) \right>,
\eeq
where $\psi_{RP}$ is the reaction plane. In the absence of dynamical $\cal{P}$--odd correlations in the emission of positive and negative particles the average (\ref{defa}) factorizes into the product of the averages of two sines 
of the emission angle, which should vanish because the system is symmetric w.r.t. the reflection around the reaction plane. The (unbroken) charge symmetry implies $a_+ a_+ = a_-a_- > 0$; because the negative and positive charges 
in the presence of charge separation effect move to the opposite sides of the reaction plane, we also expect 
$a_+a_- < 0$. Assuming the angular distribution of the produced particles to be  \cite{Kharzeev:2004ey}
\beq
{dN \over d \phi } \sim 1 + 2 a \sin \phi,
\eeq
where $\phi$ is the emission angle w.r.t. the reaction plane, the asymmetry $A$ which we have estimated is 
related to the parameter $a$ by a simple relation: $A = \pi a/4$. The magnitude of the correlation 
presented in Fig.1 thus appears comparable to the expected one. The vanishing of the asymmetry in central collisions (with angular momentum $l=0$) and its increase towards more peripheral ones 
is also qualitatively consistent with expectations. We therefore urge the continuation 
of experimental studies; we also note that 
a more direct way of observing the quark--antiquark asymmetry would be provided by the studies of baryon--antibaryon production. 

\medskip

We would like to complete the section with a cautionary remark on the accuracy of our estimates:
all theoretical formulae discussed above have been derived at zero temperature using the effective lagrangian approach. We expect that both the absolute value of the asymmetry and the real time dynamics of charge separation will  change in a more realistic calculation at $T\neq 0$. However, 
since the effect we consider is of topological nature, it should persist
at $T\neq 0$.  In addition,  the signal can be reduced if a large number of $\theta$ domains is produced in each event; however we do not think this is likely because the typical size of the domain is not much smaller than the size of the produced matter.  A more detailed treatment will be presented elsewhere; the main goal of the present letter is to demonstrate the topological nature of the charge separation effect as a manifestation 
of strong $\cal{CP}$ violation in QCD.

 \section{Possible Implications for Astrophysics and Cosmology }
  \subsection{Little Bang versus Big Bang}
$\bullet$   In the concluding section of this paper we would like to speculate on the possible consequences 
  of the charge separation phenomenon for the evolution of the Universe shortly after the Big Bang
  during the QCD phase transition at $T_0\sim 170$ MeV.
   In the axion framework, see refs \cite{axion} - \cite{review}, 
   it is generally assumed that parameter $\theta$ was  not 
    zero when the temperature of the Universe $T$ was larger than $T_0$, $T> T_0$.
      The $\theta$  angle starts to relax to zero only  at the time of 
     the QCD phase transition as a result of formation of the chiral condensate: the axion acquires its mass $m_a^2\sim m_q\< \bar{q}q\>$ while  the axion potential $V_a(\theta)$ 
    being almost flat at  $T> T_0$ 
    develops the global minimum at $\theta=0$ when $T$ slowly decreases to $T \leq T_0$ due to the expansion
    of the Universe.
      Eventually, the axion field relaxes to  the 
    minimum of the potential at $\theta=0$ such that  strong ${\cal CP}$ problem is resolved
    at   present time no matter what  the original magnitude of $\theta$
    before the QCD phase transition was. 
    This picture represents the standard resolution of strong ${\cal CP}$ problem,
    see the original papers \cite{axion},\cite{KSVZ},\cite{DFSZ},
and recent  reviews \cite{review}.
    However, we stress that while the parameter $\theta$ vanishes  at present epoch, $\theta=0$, it was of order one, $\theta \sim 1$, during the QCD phase transition. We accept this standard picture of the axion evolution for the discussions
    which follow. In different words, we treat $\theta $ as being a nonzero  
    parameter of order one      during the QCD phase transition time when the charge separation phenomena (the subject of the present work) may occur.
    The crucial ingredient of this picture  is the existence of  strong ${\cal CP}$
    violation during that time. In particular  this picture does imply that quarks and anti--quarks interacted 
    very differently at that time  in a stark contrast with the results of 
    present day  experiments  which show that ${\cal CP}$ symmetry
    is preserved with an extraordinary accuracy. This new source of very large
     ${\cal CP}$ violation  in the Universe at time when   $T\sim T_0$  
      may be a crucial missing element for solving the puzzle of  baryogenesis; we will comment on this below.
    
 $\bullet$   Another important ingredient   which  is always present when  the physics of $\theta$ angle is discussed 
is  the existence of domain walls which form during the QCD phase transition,
see Sikivie in \cite{review} for a review on the subject. The existence of these domain walls 
is a direct consequence of discrete symmetry $\theta\rightarrow\theta +2\pi n$ related to the angular nature of the $\theta$ parameter. The
axion domain wall network forms at the QCD phase transition and eventually  decays by emitting axions as well as by forming the closed surfaces, bubbles
which will  also eventually decay to axions  if nothing stops this   violent collapse.
However,  
   if the number of quarks trapped inside of a bubble  is sufficiently large, the collapse stops due to the internal Fermi pressure, see \cite{Zhitnitsky:2002qa} 
 for detail discussions. Compact objects containing large numbers of (anti)quarks can thus be formed, which could play a role of cold dark matter candidates. The {\bf separation of charges}  which may happen during this period of 
 evolution of the Universe is absolutely necessary for this scenario to work. Indeed, in this case the 
 excess  of quarks over anti quarks (or vice versa)  inside the bubble with definite baryon charge could be macroscopically large
 preventing the bubble from collapse.  The $\theta$ values  on different sides of the  axion domain wall 
 separating internal and external  regions of the bubble are different.  This feature models
 the domain which may be formed at RHIC with $\theta\neq 0$  inside the domain  and $\theta=0$ outside of it. An important difference however is that in the Early Universe the domain size is determined by the (inverse) axion mass $m_a$ and is large, whereas at RHIC a much smaller domain size is expected.
 
$\bullet$  One more crucial  element of the present work is  demonstration  that
  ${\cal CP}$ odd combination $\< \theta\vec{E}\cdot \vec{\Omega}\>\neq 0 $
   is  correlated  on the scale of the fireball size, large compared to $\Lambda_{QCD}^{-1}$.
  Similar ${\cal CP}$ odd correlation $\< \theta\vec{E}\cdot\vec{B}\>\neq 0 $ is expected to occur in the early Universe where $\vec{\Omega}$ in our calculations (relevant for 
   RHIC physics) is replaced by magnetic field $\vec{B}$ presumably induced 
   on macroscopical scales in the early Universe\footnote{ The scales  relevant for the axion
   physics   where correlation $\< \theta\vec{E}\cdot\vec{B}\>\neq 0 $ is expected to occur are the  scales at which the  $\theta$ parameter varies, i.e. on the scales of order of $m_a^{-1}$
    where $m_a$ is the axion mass, $m_a\sim (10^{-6}-10^{-3})~$eV.}, see \cite{Forbes:2000gr,KPT}.
 
 \subsection{ Separation of charges: consequences for cosmology}
 $\bullet$The objects with a large number of trapped (anti)quarks, if formed,  would be absolutely stable
 macroscopical objects  with the size of order $m_a^{-1}$. They  may serve as the cold dark matter candidates \cite{Zhitnitsky:2002qa}. 
The nuggets despite being baryonic configurations would behave like nonbaryonic dark matter objects because of tiny interaction (per unit mass) with the visible baryons, see \cite{Zhitnitsky:2006vt} for details.
 The   stability of the nugget is due  to the fact that the mass of the nugget is smaller than the mass of a collection of free separated nucleons with the same baryon charge.
 In  many aspects these nuggets are similar to Witten's strangelets~\cite{Witten:1984rs}
 with very similar phenomenology.

$\bullet$  If  the separation of matter from antimatter did indeed occur  on macroscopical scales 
  during the QCD phase transition, it  
  may lead to a number of profound consequences. In particular, we speculate that  
  two of the outstanding cosmological mysteries -- the nature of dark
matter and baryogenesis -- may be explained by the idea that dark
matter consists of dense nuggets of matter as well as antimatter.  
Due to the separation of matter from antimatter, these nuggets may form
at the same QCD phase transition where ``conventional"  baryons (neutrons and protons)
form.  This would provide a natural explanation for
the similarity of the dark matter and baryon contributions to the critical mass,  $\Omega_{DM} \approx 5\Omega_{B}$, and for
the baryon asymmetry in our part of the Universe through a charge separation mechanism driven by the
strong $\cal{CP}$  violation due to the   $\theta$ term in QCD
(see~\cite{Oaknin:2003uv} for details).  Indeed, an "excess" antimatter in this case is locked
away in antimatter nuggets; no additional baryon number violation is thus required  
to explain the observed matter/antimatter asymmetry in our part of the Universe.   

It is important  to remark here that for this mechanism to be operative there is no need for 
any fine tunings (which are an unavoidable ingredient of most
  baryogenesis scenarios):   any efficiency of the nugget formation larger than  $10^{-10}$  would be sufficient 
to explain a similarity of  two contributions, $\Omega_{DM} \sim\Omega_{B}$.
This is due to the  fact that the present day baryon density/entropy ratio $n_B/s\sim 10^{-10}$ 
is 10 orders of magnitude smaller than it was 
at the QCD phase transition epoch due to the quark-antiquark annihilations. Therefore any  quark-antiquark  asymmetry (manifesting itself through the difference in the number and/or size  of nuggets and anti-nuggets)   will survive until the freeze--out time
when  quark-antiquark  annihilation stops. A large $\cal{CP}$  violation  due to  the   $\theta$ term
at early times provides a difference of order one between number density of nuggets and anti-nuggets
such that a symmetric picture with an equal number of nuggets and anti nuggets  cannot be realized\footnote{Again, there is no fine tuning  requirements. The only requirement is the following: 
$\cal{CP}$  violation must be   sufficiently large  $> 10^{-10}$ at the time of the QCD phase transition.}. 
 Baryon charge conservation will automatically provide an appropriate  baryon number density 
  $ \Omega_{B}$ to be of order of  $\Omega_{DM} $ irrespectively of any initial conditions
  or specific details of the nugget formation mechanism as long as it is sufficiently effective, i.e. occurs with a relative probability $> 10^{-10}$.   Of course, all these attractive features of our scenario are related to the fact that both contributions originate at the same time at the same
  QCD phase transition and are made of the same constituents, quarks,  see the original paper  \cite{Oaknin:2003uv} for the details. 
One should note that in general  the 
relation $\Omega_B \sim \Omega_{DM}$, within one order of magnitude, 
between the two different contributions to $\Omega$ is extremely  difficult to 
explain in models that invoke dark matter candidates not related to the 
ordinary quark/baryon degrees of freedom.

$\bullet$ Such a ``counterintuitive" 
proposal does  not in fact contradict any of the known observational constraints on
dark matter or antimatter in our Universe due to the three main reasons \cite{Zhitnitsky:2006vt}: 

  1) The nuggets carry a huge (anti)baryon charge $|B| \approx
10^{20}$ -- $10^{33}$, so they have a macroscopic size ($m_a^{-1}\sim 10^{-3}$cm)
and  a tiny number density; 

 2) They
have   nuclear densities in the bulk, so their interaction cross-section per unit mass is small
$\sigma/M \approx 10^{-13}$ -- $10^{-9}$~cm$^2$/g. This small factor effectively
replaces a  condition on weakness of  interaction of  conventional dark matter candidates such as 
weakly interacting massive particles (WIMPs); 

3) They have a large binding energy (gap
$\Delta \approx 100$~MeV) such that baryons in the nuggets are not
available to participate in the Big Bang nucleosynthesis (BBN) at $T
\approx 1$~MeV\@.

$\bullet$ In fact, this picture may also naturally explain a number of other mysteries in astrophysics and cosmology as we will now discuss. 
Several independent observations of the Galactic core suggest
  hitherto unexplained sources of energy and have detected an excess flux of  photons across a broad 
range of energies.  If charge separation indeed took place during the 
  QCD phase transition, we may be seeing the consequences of this at present by observing the  results of rare events of annihilation between the visible matter (electrons and protons)
   and antimatter nuggets. 
   In particular, the observations which are very difficult to explain by conventional means include: 
  
1) SPI/INTEGRAL observations of the galactic centre have 
detected an excess of 511 keV gamma rays resulting 
from low momentum electron-positron annihilations. The observed intensity 
  is a mystery:   after accounting for all known positron
sources, only a small fraction of the emission may be
explained~\cite{Knodlseder:2003sv}.

2) Detection by the CHANDRA satellite of diffuse X-ray emission 
from across the galactic bulge provides a puzzling picture:  
 after subtracting the known X-ray sources  
 one finds a residual diffuse thermal X-ray emission consistent 
 with  a two-temperature plasma with 
 the hot component temperature close to  $T\simeq 8 ~$ keV. According to the analysis    \cite{Muno:2004bs}
 the  hot component 
 is very difficult to understand within the standard picture. Such a plasma
would be too hot to be bound to the galactic center. The authors of ref. \cite{Muno:2004bs} 
also remark that the energy required to sustain a  plasma of  this temperature  
corresponds to the entire kinetic energy of one supernova every 3000 yr, 
which is unreasonably high.  

3) The flux of gamma rays in the 1-20~ MeV range measured by COMPTEL represents yet
another mystery. As discussed in \cite{Strong} the best fit models for diffuse galactic 
$\gamma$ rays  fit the observed spectrum well for a very  broad range of 
energies, 20 MeV- 100 GeV. It also gives a good representation of the latitude 
distribution of the emission from the plane to the poles, and
 of the longitudinal distribution. However, the model fails to explain 
 the excess in the 1-20 MeV range
 observed by COMPTEL in the inner part of the galaxy.  
   So, what is the origin   of the sources contributing to the energy band,
1 MeV  $\leq k\leq $ 20~  MeV ?

It is quite remarkable that all these mysteries in different energy bands can be naturally 
explained if charge separation indeed took place in the Early Universe:  
an excess of the  emission we observe now would be a result of annihilation  between the visible matter 
   and the nuggets of antimatter. Dark antimatter nuggets would provide an unlimited source of positrons (e$^{+}$)   within this framework as
  suggested in~\cite{Oaknin:2004mn}.
  The resonance formation of positronium 
  between  impinging galactic electrons (e$^{-}$)  and positrons (e$^{+}$) from the DM nuggets, 
      and their subsequent decay, lead to the 511 keV line. Non-resonance direct 
      $e^+e^-\rightarrow 2\gamma$ annihilation   would produce \cite{Lawson:2007kp}
      a broad spectrum at 1 MeV  
$\leq k\leq $ 20~  MeV which is  identified with the excess  observed by COMPTEL.
   It is quite remarkable that another (also naively unrelated) puzzle, 
the  diffuse X-ray emission 
observed by CHANDRA 
 may also have a common origin with the 511 keV line and excess MeV radiation 
as argued in~\cite{McNeilForbes:2006ba}.  This emission is due to
proton (rather than electron) annihilation events releasing about $2m_{p} \approx 2$~GeV
of energy per event.  It  occurs inside the nuggets, therefore, 
the available energy 2 ~GeV energy will not be  radiated in the form of GeV photons, rather it will be  
radiated   as  X-rays in keV band~\cite{McNeilForbes:2006ba}. 
Other implications, in particular, may include
  generating the seeds of a primordial magnetic field at the QCD phase transition scale 
  \cite{KPT,Forbes:2000gr,MF}.

 $\bullet$ To conclude: several independent observations of the Galactic core suggest
    unexplained sources of energy. We speculate that the excess of photons may come 
    from rare events of annihilation of visible matter with "dark"  antimatter nuggets. 
 The charge separation phenomenon which is the subject of this paper may induce the separation of  matter and antimatter and the formation of   nuggets during the QCD phase transition if $\theta\neq 0$ at that time. 
 Charge separation effect may thus hold the key to many mysteries of the Universe, and  
 we urge the experimental studies of this phenomenon in heavy ion collisions at RHIC and LHC.

\acknowledgments The authors would like to thank Edward Shuryak  for useful discussions and comments. This work was supported by the US Department of Energy under
Contract No. DE-AC02-98CH10886 (D.K.) and by the Natural Sciences and Engineering
Research Council of Canada (A.Z.).

\section*{Appendix}
The main goal of this Appendix it to show how the eqs. (\ref{charge}),(\ref{E}) follow from the 
anomalous effective lagrangian in the presence of nonzero chemical potential and nonzero $\theta(x)$.
The corresponding physics of the anomalous effective lagrangian in the presence of $\theta$ and $\mu$  is quite rich and deep with a number of connections to different branches of physics.  
In the present paper we shall give only a short sketch of how such relations arise without much details.
We assume that $\theta(x)$ depends on coordinates and time even though for the purposes of the present paper we 
do not need to assume the existence of the axion. The main reason 
for that assumption is that we expect that a small domain   with $\theta\neq 0$ can be formed 
at RHIC. This region  will be treated as a macroscopically large region
inside of a much larger region (representing our  world) with $\theta=0$.
The $\theta(x)$ will be treated as a classical background external field without 
dynamical kinetic term. 
Nontrivial boundary conditions between the regions of $\theta=0$ and $\theta\neq 0$
will play the crucial role in our discussions.
The corresponding technique with nonzero $\mu$ was developed in \cite{SZ} and generalized for the axion field (nonzero $\theta (x)$) in~\cite{MZ}.

We limit ourselves only to one term 
in the anomalous effective lagrangian which plays the crucial role for
the present study, see \cite{SZ} and \cite{MZ} for details:
  \beq 
\label{1}
 L_{\theta\gamma V} = 
- N_c\sum_f \frac{e_f \mu_f }{4 \pi^2N_f}\cdot    \eps\dm \theta (\d_{\lambda}V_{\nu} ) A_{\sigma},
 \eeq
  where  $V_{\mu} $ is a fictitious vector gauge
field  introduced in \cite{SZ}. Then the coupling of quarks to
$V_{\mu}$ and to the usual electromagnetic gauge field $A_{\mu}$ are almost identical; 
the only difference is in the coupling constants.  
This new term (\ref{1}) leads to a few very unusual phenomena. First, it leads to the separation of charge.
Indeed, by definition, the local charge density  is defined as
\beq
\label{2}
J^{ind}_0=\frac{\delta  L_{\theta\gamma V}}{\delta A_0}=
N_c\sum_f \frac{e_f \mu_f }{4N_f \pi^2}\cdot    \epsilon^{ijk}\d_i \theta (\d_{j}V_{k} ) =
  N_c\sum_f \frac{e_f \mu_f }{2 N_f \pi^2}\cdot      \left(\vec{\nabla}\theta\cdot \vec{\Omega}\right),
 \eeq
where $ \vec{\Omega}$ is defined as $ 2\epsilon_{ijk} \Omega_k= (\d_i V_j-  \d_{j}V_{i} )$ and is nothing but the angular velocity of the rotating system. The result (\ref{2})  is very suggestive 
and implies that there is a charge separation phenomenon in the presence of $\theta$ in 
the rotating system  along the vector of the rotation $ \vec{\Omega}$.
Indeed,  assuming that  $ \vec{\Omega}$ is directed along $z$ we arrive to the following 
expression for the density charge ${Q}/{\Sigma_{xy}} $ ( per surface area $\Sigma_{xy}$ in the plane perpendicular to $\Omega_z$)    accumulated at $z=\pm L/2$ where there is a transition between  induced $\theta$ state and vacuum\footnote{here and in what follows we assume a cylindrical geometry defined by the 
length $L$ along $z$ and area $\Sigma_{xy}$},
\beq
\label{3}
\sigma_{xy}\equiv \frac{Q}{\Sigma_{xy}}=\int^{L/2}_{0}  dz
J^{ind}_0 = 
  -N_c\sum_f \frac{e_f \mu_f  }{ 2\pi^2}\cdot    \Omega_z \frac{\theta}{N_f}, 
   \eeq
 where $\theta$ is a value of $\theta(x=0)$ inside the bubble for a given event while $\theta=0$ 
 in the vacuum, outside the region of interest.
A charge with the opposite sign will be induced at $z=-L/2$.
This is precisely eq. (\ref{charge}) used in the text.

Let us rewrite eq. (\ref{1}) in the following way,
\beq
\label{4}
L=\frac{1}{2} \vec{E}^2  +
  N_c\sum_f \frac{e_f \mu_f}{2N_f \pi^2}\cdot    \theta  \left(\vec{E}\cdot \vec{\Omega}\right),
 \eeq
where we added the standard  kinetic term $ \frac{1}{2} \vec{E}^2  $.
Without the anomalous term the ground state corresponds to $\<\vec{E}\>=0$.
However, in the presence of the  $\theta$ term the minimum of the free energy corresponds 
to the state where $\<\vec{E}\>\neq 0$. Indeed, minimization of (\ref{4}) with respect to $\vec{E}$ gives,
\beq
\label{5}
\frac{\delta L}{\delta E}= \vec{E}+
N_c  \sum_f \frac{e_f \mu_f}{ 2\pi^2}\cdot      \left(\frac{\theta}{N_f} \right)  \cdot  \vec{\Omega} =0,
 \eeq
which  is precisely eq. (\ref{E}) used in the text.
    As is known, the electric field $E_z$  between two 
infinitely large  charged plates   with charge density $\sigma_{xy}$ is exactly equal 
$E_z=\sigma_{xy}$. Our equations (\ref{3}) and (\ref{5})
satisfy this famous equation for an infinitely large capacitor even though they were derived in a very different way.

One more remark on eq. (\ref{5}). An additional term in the lagrangian
(\ref{4})   is a total derivative, and therefore does not change the equations of motion.
However, in order to quantize 
the system  (\ref{4}) in canonical way,  we have to express everything in terms  of the 
  generalized momentum $ {\pi_i}=
\frac{\delta L}{\delta \dot{A_i}}$ 
which is defined as 
\beq
\label{pi}
 {\pi_i}=
\frac{\delta L}{\delta \dot{A_i}}=  {E_i}+
  N_c  \sum_f \frac{e_f \mu_f}{ 2\pi^2}\cdot      \left(\frac{\theta}{N_f} \right)  \cdot  {\Omega_i} 
 \eeq
 In the ground state we require that  $\<{\pi_i}\>=0$ which is precisely the condition (\ref{5}) derived 
 earlier. In different words, while the $\theta$ term is a total derivative and does not change the equations of motion, it does influence the physics (separation of charges, induced electric field)
 in the presence of topological background field, $   \Omega_k\sim \epsilon_{ijk}(\d_i V_j-  \d_{j}V_{i}) 
 \neq 0$.

    In order to integrate over the surface $d\Sigma_{xy}$ 
 we have to understand the way how the rotation is quantized.
 The answer lies in the definition of the fictitious 
field $V_{\nu}\equiv v_{\nu} $ which is  
the local $4-$ velocity  of matter with chemical potential $\mu_f$, see \cite{SZ}\cite{MZ}.
Then the coupling of quarks to
$V_{\mu}$ and to the usual electromagnetic gauge field $A_{\mu}$
takes the form: \beq {\cal L} = \sum_f (\mu_f V_{\nu} - e_f
A_{\nu}) \bar{\psi}_f \gamma_{\nu} \psi_f \eeq
With this definition 
magnetic flux quantization as well as rotation 
quantization for a single species  are very similar and take the form,
\bea
\label{6}
 \int e  \vec{B}\cdot  \ d\vec{\Sigma} =2\pi l 
 \\
  \int \mu  \vec{\Omega} \cdot \ d\vec{\Sigma} =2\pi l
  \eea

\end{document}